\documentclass[useAMS,usenatbib]{mn2e}

\usepackage{graphicx,amssymb}
\usepackage[normalem]{ulem}
\usepackage[usenames,dvipsnames]{xcolor}
\usepackage{soul}

\newcommand{\rev}{ }

\title[Planet formation around $6-8M_{\odot}$ stars]
{Constraining planet formation around 6$M_{\odot}$-8$M_{\odot}$ stars}

\author[Veras et al.]{Dimitri Veras$^{1,2}$\thanks{E-mail: d.veras@warwick.ac.uk}\thanks{STFC Ernest Rutherford Fellow},
Pier-Emmanuel Tremblay$^{1}$, J.J. Hermes$^{3}$,
\newauthor 
Catriona H. McDonald$^{1,2}$, Grant M. Kennedy$^{1,2}$\thanks{Royal Society University Research Fellow}, Farzana Meru$^{1,2}$\thanks{Royal Society Dorothy Hodgkin Fellow}, 
\newauthor
Boris T. G\"{a}nsicke$^{1,2}$
\\
$^{1}$Centre for Exoplanets and Habitability, University of Warwick, Coventry CV4 7AL, UK
\\
$^{2}$Department of Physics, University of Warwick, Coventry CV4 7AL, UK
\\
$^{3}$Department of Astronomy, Boston University, 725 Commonwealth Ave., Boston, MA 02215, USA
}

\pubyear{2019}

\begin{document}
\label{firstpage}
\pagerange{\pageref{firstpage}--\pageref{lastpage}}
\maketitle

\begin{abstract}
Identifying planets around O-type and B-type stars is inherently difficult; the most massive known planet host has a mass of only about $3M_{\odot}$. However, planetary systems which survive the transformation of their host stars into white dwarfs can be detected via photospheric trace metals, circumstellar dusty and gaseous discs, and transits of planetary debris crossing our line-of-sight. These signatures offer the potential to explore the efficiency of planet formation for host stars with masses up to the core-collapse boundary at $\approx 8M_{\odot}$, a mass regime rarely investigated in planet formation theory. Here, we establish limits on where both major and minor planets must reside around $\approx 6M_{\odot}-8M_{\odot}$ stars in order to survive into the white dwarf phase. For this mass range, we find that intact terrestrial or giant planets need to leave the main sequence beyond approximate minimum star-planet separations of respectively about 3 and 6 au. In these systems, rubble pile minor planets of radii 10, 1.0, and 0.1 km would have been shorn apart by giant branch radiative YORP spin-up if they formed and remained within, respectively, tens, hundreds and thousands of au. These boundary values would help distinguish the nature of the progenitor of metal-pollution in white dwarf atmospheres. We find that planet formation around the highest mass white dwarf progenitors may be feasible, and hence encourage both dedicated planet formation investigations for these systems and spectroscopic analyses of the highest mass white dwarfs. 
\end{abstract}


\section{Introduction}

From the first detection of exoplanetary material one century ago (\citealt* {vanmaanen1917, vanmaanen1919}; but recognized as such only much later) to the pioneering discoveries of exoplanets over the last three decades \citep{cametal1988,latetal1989,wolfra1992,hatcoc1993,wolszczan1994,mayque1995,marbut1996}, we are starting to piece together an understanding of planetary formation, evolution and fate. Major and minor planets, including asteroids, comets and planetary debris, have now been observed orbiting brown dwarfs, M-type to B-type main-sequence stars, subgiant and red giant branch stars, white dwarfs, and pulsars. However, we still lack definitive detections in systems with O-type stars, asymptotic giant branch stars, and black holes\footnote{\cite{keretal2016} reported a so far unconfirmed candidate planet orbiting an asymptotic giant branch star.}.

On 21 Nov 2019, the NASA Exoplanet Archive listed as ``confirmed" 4,099 major planets which have been discovered orbiting main-sequence, subgiant and giant branch stars. From that population, the most massive known planet host with a well-constrained (within 10 per cent) mass above $3.0M_{\odot}$ is {\it o} UMa b, whose host star mass is $3.09 \pm 0.07M_{\odot}$ 
\citep{satetal2012}. This detection cutoff at about $3.0M_{\odot}$ is not sharp, but rather represents a tail reflecting a decreasing number of discoveries as a function of stellar mass (see Fig. 5 of \citealt*{refetal2015}, Fig. 9 of \citealt*{gheetal2018} and Fig. 3 of \citealt*{gruetal2019}).

This upper bound results from some combination of observational limitations and restrictions on where and how planets can form and evolve around stars more massive than the Sun \citep{kenken2008}, and does not change depending on stellar classification \citep{lloyd2011} nor whether asteroseismological constraints are taken into account \citep{cametal2017,noretal2017,steetal2017}.

The formation of planets around (low-mass) $\le 3M_{\odot}$ stars represents one of the most extensively studied aspects of exoplanetary science. The two oldest formation theories are gravitational instability \citep{kuiper1951,cameron1978,boss1997} (see \citealt*{duretal2007}, \citealt*{heletal2014}, \citealt*{kralod2016} and \citealt*{rice2016} for reviews), and core accretion \citep{safzvj1969,golwar1973,percam1974,harris1978,mizuno1980} (see \citealt*{pollack1984} and \citealt*{lisste2007} for reviews). More recently, important physical processes such as the streaming instability \citep{yougoo2005,johetal2007} and pebble accretion \citep{lamjoh2012} have provided alternatives or enhancements to the traditional formulations. Each theory, when combined with post-formation dynamical evolutionary pathways, succeeds in reproducing observed properties of some but not all of the known exoplanets. 

All these formation processes require the presence of a circumstellar, protoplanetary disc, and operate in different regions of the disc on different timescales. A disc must survive long enough and be massive enough for a planet to form. 

Unfortunately, like observations of main-sequence exoplanet host stars themselves, observations of protoplanetary discs are restricted by the rarity of massive host stars and the short lifetimes of their discs. Early on in the life of a stellar birth cluster, circumstellar discs have been observed to contain a wide variety of masses, up to host masses of many tens of solar masses \citep{foretal2016,ileetal2018,janetal2019,sanetal2019}. Young, intermediate stars such as Herbig Ae/Be stars are commonly seen to host circumstellar discs, which are almost certainly sites of planet formation (much as T Tauri stars are precursors to planet-hosts like the Sun). Several of these stars have masses greater than $3M_{\odot}$ \citep{heretal2005}, suggesting that a fertile circumstellar environment could exist for high host-star mass planet formation.

Here, we present another source of motivation to study planet formation around high-mass ($6M_{\odot} - 8M_{\odot}$) stars: the end state of these systems. As outlined in Section 2, current observations of white dwarfs with exoplanetary material expand the traditional main-sequence host-star mass range, and future observations could expand this range even further. In fact, the vast majority of $\gtrsim 3M_{\odot}$ stars ever formed and which now reside within a local volume of hundreds of parsecs are cooling white dwarfs. 

Based on these prospects, we compute survival limits for the bounding case of the highest main-sequence host star masses which would not trigger a core-collapse supernova\footnote{\cite{wadetal2019} considered planet formation around supermassive black holes which are just a few parsecs away from active galactic nuclei. However, the prospects of detecting planetary systems outside of the Milky Way Galaxy are currently remote, except perhaps for the Large Magellanic Cloud (which probably does not contain a central black hole; \citealt*{covetal2000,ingetal2009,lunetal2015,mropol2018}).} ($\approx 6M_{\odot}-8M_{\odot}$) but instead become white dwarfs (which could eventually harbour observed metal pollution). In Sections 3 and 4 respectively we compute survival limits for both major and minor planets around these massive progenitors. We discuss our results in Section 5 and conclude in Section 6, hoping to motivate future dedicated investigations of planet formation around high-mass stars.

\section{White dwarf planet hosts}

After both major and minor planets are formed and dynamically settle, the remainder of main-sequence evolution is thought to remain relatively quiescent. For example, in our solar system, over the last 4 Gyr or so, the orbits of the eight major planets have not varied enough to incite mutual scattering events. This quiet situation is very likely to continue until the end of the Sun's main-sequence lifetime about 6 Gyr from now \citep{lasgas2009}.

Subsequently, when the Sun ascends the giant branch phases, major changes (described in detail in Sections 3 and 4) will ensue. The surrounding bodies which survive these changes will eventually orbit a white dwarf, a stellar ember that is dense enough to stratify any accreted material into its constituent chemical elements \citep{schatzman1945,schatzman1947}. This property enables the detection of exoplanetary metals, because predominantly white dwarf atmospheres would otherwise be composed of only hydrogen and/or helium.

\subsection{Origin of exoplanetary material}

These metallic debris seen in the atmospheres of single white dwarfs do not necessarily arise from major planets. In fact, signatures of individual minor planets have now been found around three different white dwarfs: WD 1145+017 \citep{vanetal2015}, SDSS J1228+1040 \citep{manetal2019} and ZTF J0139+5245 \citep{vanetal2019}.

These discoveries have corroborated observations of metal-polluted white dwarf photospheres. On both chemical and dynamical grounds, most of the atmospheric debris likely originates from exo-asteroids \citep{ganetal2012,juryou2014,veras2016a,haretal2018,holetal2018,doyetal2019,swaetal2019}, exo-moons \citep{payetal2016,payetal2017} or their fragments from destructive giant branch radiation \citep{veretal2014a,versch2019}. Minor planets, or their fragments, can avoid being engulfed into the star during the giant branch phases only at distances beyond several or many au \citep[]{musvil2012,madetal2016,galetal2017,raoetal2018,sunetal2018}. Therefore, at least one (surviving) major planet must exist in these systems to perturb the minor bodies into the disruption region around the white dwarf from au-scale distances, as argued extensively in \cite{veretal2018a}. 

What if the white dwarf is currently observed to have a binary stellar companion \citep{pyretal2012,wiletal2019}? Then, debris in the white dwarf atmosphere could originate from the winds of its partner rather than a planetary system. However, if the companion is separated beyond a critical distance, then the accreted mass from the wind would be too small to explain the observations (and hence must arise from a planetary system). This critical distance is on the order of few au \citep{debes2006,veretal2018a}, well within the binary separation for the majority of known polluted white dwarfs in binaries. 

What if, instead, a polluted white dwarf {\it used to} have a binary stellar companion which underwent a merger\footnote{\citet{Cheng2019} have recently suggested that 20 $\pm$  6 per cent of high-mass white dwarfs ($1.08M_{\odot} - 1.23M_{\odot}$) originate from double white dwarf mergers \citep{toonen2012,shen2012,garcia-berro2012}. In addition, another significant fraction are likely merger byproducts merged when at least one of the stellar components was not a white dwarf \citep{toonen2017,temmink2019}.}? Could the atmospheric debris then result from the merger event? The answer is no, because any metal mixing resulting from the merger would sink to the core very quickly, on a timescale which is 3-11 orders of magnitude shorter than the cooling age \citep{koester2009,dufour2010,baubil2018,baubil2019,cunetal2019}. Hence, no chemical signatures from the merger would be currently observable.

In summary, regardless if a white dwarf has or had binary companions, {\it both major and minor planets should exist in metal-polluted systems without stellar binary companions
and whose main-sequence progenitor masses were $6M_{\odot}-8M_{\odot}$}. So far major planets in such high mass systems have not been observed\footnote{However, see \cite{gaensicke2019} for robust evidence of a major ice giant planet orbiting a white dwarf whose progenitor mass was $1.0M_{\odot} - 1.6M_{\odot}$; \cite{verful2020} has suggested that this giant planet is a partially destroyed and highly inflated ``Super-Puff".}. However, the statistics of high-mass metal rich white dwarfs are still poorly understood, owing to the steepness of the initial mass function and their faintness, and therefore their rarity \textcolor{black}{\citep{tremblay2016}}.

\begin{table*}
 \centering
 \begin{minipage}{170mm}
 \centering
  \caption{Some stellar evolution properties for given zero-age-main-sequence (ZAMS) mass and metal mass fraction (given through the quantity $Z$) tracks based on the {\tt SSE} code. ``AGB'' is an abbreviation for asymptotic giant branch. For the metal-poor $Z=0.0001$ case, the $7M_{\odot}$ and $8M_{\odot}$ stars become neutron stars, and hence we do not consider them further.}
  \label{stelprop}
  \begin{tabular}{@{}cccccccc@{}}
  \hline
  ZAMS mass     & $Z$     & white dwarf mass    & main sequence duration     & AGB duration   & Max $R$    & Max $L$         & Max $T$ \\
  ($M_{\odot}$) &             & ($M_{\odot}$)  &  (Myr)      & (Myr)  & (au)    &  ($L_{\odot}$)               & (K) \\[2pt]
 \hline
     $6.00$          & 0.02    &   1.14                  & 66.0         & 0.56    &  5.94   &  $6.16 \times 10^4$    & $4.12 \times 10^5$    \\[2pt]
 \hline
    $6.00$          & 0.0001 &  1.37                   & 60.0        & 0.75     &  4.29  & $1.54 \times 10^5$     & $6.93 \times 10^5$  \\[2pt]
 \hline
    $7.00$          & 0.02     & 1.29                     &  48.9       & 0.51     & 6.73   & $7.67 \times 10^4$      &  $5.07 \times 10^5$   \\[2pt]
 \hline
    $8.00$          & 0.02     &  1.44                    &   37.2      &  0.49    & 7.39   &  $9.16 \times 10^4$     &  $1.11 \times 10^6$  \\[2pt]
 \hline
\end{tabular}
\end{minipage}
\end{table*}

\subsection{The most massive white dwarf host}

Unlike for main-sequence exoplanetary systems, white dwarf planetary systems are not yet summarized with a publicly-available database. Rather, this data is contained within different published studies \citep{dufetal2007,kleetal2013,genetal2015,kepetal2015,kepetal2016,holetal2017,genetal2019,coutuetal2019}. 

Nevertheless, we can identify the most massive polluted white dwarf host based from individual observing campaigns. The homogeneous sample of bright, hydrogen-atmosphere white dwarfs studied in \cite{koeetal2014} reveals that the metal polluted WD\,1038+633 has a robust mass determination both from Balmer line spectroscopy ($M_{\rm WD}$ = 0.90~$\pm$ 0.01~$M_{\odot}$; \citealt{tremblay2011}) and from \textit{Gaia} photometry and astrometry ($M_{\rm WD}$ = 0.91 $\pm$ 0.01 $M_{\odot}$; \citealt{genetal2019}). The progenitor mass is more uncertain, but \citet{koeetal2014} estimate a value of $\approx$ 4~$M_{\odot}$. 

In a different investigation, \citet{coutuetal2019} analysed a sample of helium-atmosphere white dwarfs with metal pollution mostly drawn from the Sloan Digital Sky Survey (SDSS). Their figure 10 demonstrates that for a generous upper limit of 10\% precision on \textit{Gaia} parallax, the most massive currently known progenitor of a metal polluted white dwarf is similarly $\approx$ 4~$M_{\odot}$, albeit with larger error bars on the white dwarf mass. 

In either case, the most massive currently observed metal-polluted white dwarf contains a progenitor mass of about $4M_{\odot}$. In Section 5, we will elaborate on the exciting and imminent increase in the number of polluted white dwarfs for which spectroscopy will be obtained, and the prospects for finding even higher mass hosts.

\subsection{Main-sequence progenitor masses}

Extrapolating measured white dwarf masses to main-sequence values requires the application of an
initial-to-final-mass relation. This relation, however, is poorly constrained in the $6M_{\odot}-8M_{\odot}$ regime. Observationally, Figure 5 of \cite{cumetal2018} reveals that white dwarf masses higher than about $1.05M_{\odot}$ 
plausibly correspond to main-sequence masses of about $6M_{\odot}-8M_{\odot}$, with large uncertainties; the lower mass boundary for a core-collapse supernova to occur is $8M_{\odot} \pm 1M_{\odot}$ \citep{smartt2009}.

Plots for the maximum envelope radius, phase durations, main sequence age, and mass loss 
for $6-8 M_{\odot}$ stars based on the {\tt SSE} stellar evolution code \citep{huretal2000} have already 
all been presented in \cite{veretal2013a} and we do not repeat them here. Instead, we reiterate
the most important of those features in Table \ref{stelprop}. Variations in these numbers which would arise
from the application of other stellar codes are small enough to not qualitatively affect our results,
especially given the uncertainty in the initial-to-final-mass relation.

\subsection{Short main-sequence lifetimes}

One set of values from Table \ref{stelprop} which deserves special attention is the duration of the main sequence phase (35-70 Myr).  Hence, subsequent to protoplanetary disc dissipation, for $6M_{\odot}-8M_{\odot}$ systems, only a few to several tens of Myr would remain
before the host star leaves the main sequence. This period of time {\rev might}
have been too short for the giant planets in our own solar system to 
dynamically settle \citep[e.g.][]{tsietal2005,moretal2018}. Nevertheless, significant gravitational scattering events may occur earlier in other systems, before the end of the main sequence phase. 

Even during the main-sequence phase of these systems, planets are subject to other forces which are often neglected in other exoplanetary studies. In the most extreme case of a $8 M_{\odot}$ host star, the star will actually lose 3.4 per cent of its mass by the end of the main sequence. In contrast, the Sun will lose no more than 0.06 per cent of its mass before the end of the main sequence \citep{verwya2012,airusm2016}. Further, how planet formation within protoplanetary discs is affected by such a rapidly changing
massive star remains uncertain.

\subsection{White dwarf disc lifetimes}

Another important aspect of white dwarf planetary system observations is their age, because that parameter helps constrain dynamical history. The ``cooling age'' in particular refers to the time since the white dwarf was born. Metal polluted white dwarfs with cooling ages longer than about 100 Myr have metal sinking timescales in the atmosphere that range from just a few weeks to a few Myr \citep{koester2009}\footnote{The lower end of the range corresponds to warm hydrogen-atmosphere (DA class) white dwarfs.}.

Such a short sinking timescale relative to the cooling age implies that metal pollution arises from a steady accretion stream originating in a ring- or disc-like planetary debris structure; these structures have now been observed in over
40 systems \citep[e.g.][]{zucbec1987,graetal1990,gaeetal2006,farihi2016,denetal2018} and many others are likely hidden from view \citep{bonetal2017}. Supporting the accretion stream theory is that white dwarfs, being Earth-sized, 
represent targets which are too small for direct hits by minor planets scattered from distances of many au \citep{broetal2017}. 

The lifetime of those discs provides a hint 
as to when the surviving major planet (or planets) perturbed a minor planet (or its fragments) towards the white dwarf. Without considering any disc replenishment mechanisms \citep{griver2019,makver2019}, the lifetime of these discs is estimated to be $\sim 10^4 - 10^6$ yr \citep{giretal2012}. Because
these timescales are still much smaller than a typical cooling age of at least 100 Myr, in this case the major planet did not trigger the pollution of the star immediately after it became a white dwarf,
and the planet must have survived for at least $100$ Myr of white dwarf cooling.

\section{Major planet survival}

Theoretical efforts to link main-sequence and white dwarf planetary systems in the highest-mass regimes have been sparse.  \cite{veretal2011,veretal2013a} and \cite{voyetal2013}
did previously consider the evolution of major planets around $6M_{\odot}-8M_{\odot}$ host stars. Since 2013, however, the focus has shifted towards 
much lower stellar masses (typically $3M_{\odot}$ and below) because those were the only ones which matched 
the mounting number of observations over this period 
(see in particular the match between the results of \citealt*{musetal2018} and \citealt*{holetal2017,holetal2018}). In this section, we provide a significantly updated assessment of the basic survival considerations from \cite{veretal2013a} and \cite{voyetal2013}.

As argued earlier, in order for the eventual white dwarf to be polluted at a cooling age of hundreds of Myr, 
at least one major planet needs to survive until the white dwarf phase, and then for
at least another approximately 100 Myr. Subsequently the major planet would kick a minor planet or debris
towards the white dwarf (with the travel time being less than 1 Myr). A disc is then formed,
and that disc would accrete onto the star on a timescale less than about 1 Myr.

Now we detail how the major planet can survive until the white dwarf phase.

\subsection{Orbital expansion due to mass loss}

During the giant branch phases, orbits will expand due to stellar mass loss
(\citealt*{omarov1962}; \citealt*{hadjidemetriou1963}; see Section 4 of \citealt*{veras2016a} for a historical
summary). For isotropic mass loss, an ``adiabatic'' regime
may be defined where the orbital eccentricity remains nearly constant as the semimajor
axis is increased \citep{veretal2011}. In this regime, the semimajor axis increases 
inversely to the mass which remains, such that, for example,
a 75 per cent mass loss quadruples the semimajor axis.
The boundary of the adiabatic regime scales with stellar mass. 
However, Figs. 14-15 of \cite{veretal2011} 
illustrate that despite the high mass loss
rate from $6-8M_{\odot}$ stars, the resulting
change in eccentricity for orbiting bodies within $100$ au remains negligible
(and in the adiabatic regime). 

If the mass loss is instead anisotropic, then the character of the orbital
expansion changes \citep{veretal2013b,doskal2016a,doskal2016b}.
However, the potential anisotropy of the mass loss from $6-8M_{\odot}$ stars
is not well-constrained, and evidence from planetary nebulae remain
in the realm of binary stars \citep[e.g.][]{hiletal2016}. Only significant sustained anisotropy 
would change the orbital evolution of a planet within 100 au from its
predicted isotropic adiabatic value \citep{veretal2013b}.

In summary, given the mass values in Table \ref{stelprop}, the semimajor axis 
of an orbiting planet would likely increase by a factor of 4.4--5.6. For the lower mass stars
usually considered in post-main-sequence planetary science studies, this factor is
instead about 2--4 \citep[Fig. 3 of][]{veras2016a}.

\subsection{Orbital engulfment due to tidal interactions}

As the star is losing mass and the planet's orbit is expanding, the star is expanding
as well. In fact, the stellar envelope may expand quickly enough and close enough to the planet
to draw it inside. Many investigations
have computed the critical engulfment distance (technically the minimum star-planet separation
on the main sequence which leads to engulfment on the giant branches) for the red giant branch phase
\citep{villiv2009,kunetal2011,viletal2014,galetal2017,raoetal2018,sunetal2018},
the asymptotic giant branch phase \citep{musvil2012}, or both phases
\citep{adablo2013,norspi2013,madetal2016}.

The approaches and prescriptions in the papers listed above differ. Here
we seek just a rough estimate. Further, for $6-8 M_{\odot}$ stars,
the red giant branch phase is negligibly short. Consequently, the extremes of luminosity, temperature and radius all occur close to, but not precisely at, the tip of the asymptotic giant
branch phase. 

Hence, we focus on this latter phase only.
\cite{musvil2012} carried out detailed numerical simulations which determined
the critical engulfment distance of different types of planets (different masses and compositions) around asymptotic
giant branch stars whose main-sequence progenitor masses went up to $5M_{\odot}$.
They used the equilibrium tidal model of \cite{zahn1977} for their formalism. \cite{adablo2013}
instead took a different approach, and derived an analytical formula for the critical engulfment
distance, $d_{\rm eng}$, as a function of several free parameters. Their prescription includes a correction
for pulsations during the asymptotic giant branch phase. Neither study adopted stellar metallicity
as a free parameter\footnote{More massive progenitor stars should have higher metallicities because they must be younger if they are evolved from single stars.}.

Let us assume the extreme quiescent case of a planet forming on a circular orbit around 
stars more massive than $5M_{\odot}$. Extrapolating Fig. 7 of \cite{musvil2012} yields
the following crude estimates for Jovian, Neptunian and terrestrial planets:

\begin{equation}
d_{\rm eng}^{\rm (Jovian)} = 0.48 \ {\rm au} \left( \frac{M_{\star}}{M_{\odot}} \right) + 2.6 \ {\rm au} 
\end{equation}

\[
\ \ \ \ \ \ \ \ \ \ \ \, = \left\lbrace 5.5, 6.0, 6.4 \right\rbrace \ {\rm au},
\]

\begin{equation}
d_{\rm eng}^{\rm (Neptunian)} = 0.33 \ {\rm au} \left( \frac{M_{\star}}{M_{\odot}} \right) + 1.8 \ {\rm au} 
\end{equation}

\[
\ \ \ \ \ \ \ \ \ \ \ \ \ \, = \left\lbrace 3.8, 4.1, 4.4 \right\rbrace \ {\rm au},
\]

\begin{equation}
d_{\rm eng}^{\rm (Terrestrial)} = 0.28 \ {\rm au} \left( \frac{M_{\star}}{M_{\odot}} \right) + 1.42 \ {\rm au} 
\end{equation}

\[
\ \ \ \ \ \ \ \ \ \ \ \ \ \ \ \, = \left\lbrace 3.1, 3.4, 3.6 \right\rbrace \ {\rm au},
\]

\noindent{}where the evaluations above correspond to ZAMS masses of
$M_{\star} = \left\lbrace 6,7,8\right\rbrace M_{\odot}$.   

Equations 32 and 38 of \cite{adablo2013} instead give

\[
\frac{d_{\rm eng}}{2 \ {\rm au} } 
\approx \left[  
\mathcal{P}
\left( \frac{q+1}{q + \beta  - p - 1} \right)
\left( \frac{\tau}{0.1 \ {\rm Myr}} \right)
\left( \frac{M_{\rm planet}}{M_{\rm Jupiter}} \right)
\right]^{2/19}
\]

\begin{equation}
\ \ \ \ \ \ \ \times 
\left( \frac{M_{\star}}{M_{\odot}}  \right)^{-1/19}
\left( \frac{R_{\star}}{1 \ {\rm au}}  \right)^{16/19}
.
\label{tidcrit}
\end{equation}

\noindent{}In equation (\ref{tidcrit}), $\tau$ represents a mass loss timescale; here we use the
duration of the asymptotic giant branch phase (see Table \ref{stelprop}). The coefficient
$\mathcal{P} > 1$ represents the enhancement factor due to the presence of pulsations.
The indices
$q$, $p$ and $\beta$ respectively represent the radial dissipation index, the mass loss
index and the mass dissipation index. Although these values are free parameters, 
 \cite{adablo2013} stated that the critical value of the tidal dissipation parameter,
which is a component of equation (\ref{tidcrit}), is accurate to within 50 per cent
across the entire parameter space.

Our own numerical investigation reveals that the leading term with the indices
varies by less than 20 per cent across all plausible values of $q$, $p$ and $\beta$.
Hence, we simply adopt $q=7$, $\beta = 2$, and $p=0$, such that the leading term is unity.
We also take both $M_{\star}$ and $R_{\star}$ to represent the mass and radius of at the star 
at the start of the asymptotic giant branch phase.  Then for ZAMS masses of
$M_{\star} = \left\lbrace 6,7,8\right\rbrace M_{\odot}$, we obtain

\begin{equation}
d_{\rm eng}^{\rm (Jovian)} = \left\lbrace 4.0, 4.3, 4.5 \right\rbrace\mathcal{P} \ {\rm au}, 
\end{equation}

\begin{equation}
d_{\rm eng}^{\rm (Neptunian)} = \left\lbrace 3.0, 3.2, 3.3 \right\rbrace\mathcal{P} \ {\rm au}, 
\end{equation}

\begin{equation}
d_{\rm eng}^{\rm (Terrestrial)} = \left\lbrace 2.2, 2.4, 2.5 \right\rbrace\mathcal{P} \ {\rm au}. 
\end{equation}

\begin{figure}
\includegraphics[width=8cm]{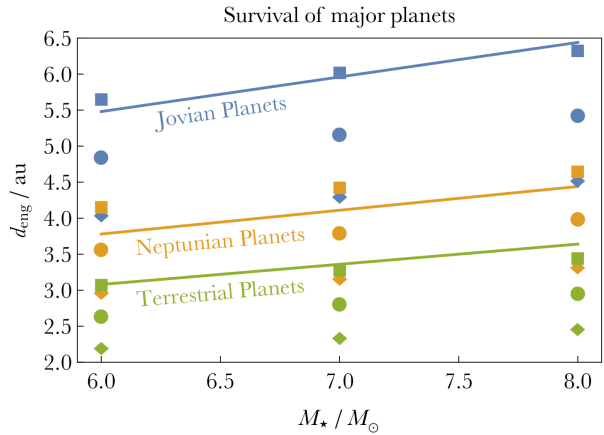}
\caption{
Minimum main-sequence star-planet distances ($d_{\rm eng}$) for which a major 
planet will avoid engulfment in the asymptotic giant branch stellar envelope and 
survive until the white dwarf phase. The solid lines are simple linear extrapolations
from Mustill \& Villaver (2012) and the symbols are applications of Adams \& Bloch (2013). The diamonds
represent the (unrealistic) case when no pulsations along the asymptotic giant branch phase
are taken into account. The circles and squares assume a pulsation enhancement factor of
20 and 40 per cent, respectively. The stars are assumed to have solar metallicity.
}
\label{major}
\end{figure}

Figure 4 of \cite{adablo2013} illustrates that the pulsation enhancement factor 
$\mathcal{P}$ can change the engulfment
boundary by several tens of per cent. Hence, the values of $d_{\rm eng}$ from the two different
sets of assumptions and extrapolations employed here may be roughly consistent.  We show
the correspondence in Fig. \ref{major}.

\subsection{In-spiral inside stellar envelope}

A planet engulfed in a stellar envelope is not necessarily immediately destroyed. The possibility exists
that the planet may survive the resulting in-spiral. If so, the envelope -- which is initially puffed up
by the planet ingestion \citep{staetal2016} -- would have to dissipate before
the planet becomes compressed and flattened by ram pressure \citep{jiaspr2018}.
Planets themselves are typically too small to unbind the envelope \citep{norbla2006},
and so survival would depend on a slow in-spiral.

By using orbital decay power scalings, \cite{macetal2018} compute, in their Figure 4, the in-spiral timescale in terms of orbits
for a Jovian planet across parts of the Hertszprung-Russell diagram. In every case the number of in-spiral
orbits is less than, and usually much less than, $10^4$ orbits. At most then, the in-spiral time represents
about 10 per cent of the duration of the asymptotic giant branch phase. 

The in-spiral time is hence effectively negligible. This conclusion, however, is extrapolated from
just the handful of papers above, which focus on the highest mass planets and stars with masses lower
than those we consider here. More detailed modelling across a wider parameter space would provide
a necessary complement to the tidal studies cited in Section 3.2.

Other related investigations have focused on the consequences for the parent star.
\cite{neltau1998} illustrated that engulfment could lead to an under-massive white
dwarf. 
Potential lithium enhancement due to planet engulfment \citep{caretal2012,caretal2013,steetal2019}
would no longer be observable by the time the star becomes a white dwarf.
However, rotational spin-up of the giant branch star due to engulfment \citep{massarotti2008,steetal2019}
might leave a small but lasting residue, because the spin period of the eventual white dwarf would affect
tidal interactions with orbiting planets \citep{veretal2019a,verwol2019,verful2019,verful2020}.

The residue perhaps also is manifest in the resulting change of magnetic field strength \textcolor{black}{\citep{farihi2011,kissin2015}}. The \textit{Gaia}-SDSS spectroscopic white dwarf catalog in \cite{genetal2019} reveals that up to 50\% of white dwarfs whose progenitors had masses of $6-8 M_{\odot}$ have magnetic fields above the typical 1 MG detection limit \citep{kepler2013}. This fraction could, however, represent a signature of the large fraction of massive white dwarfs that are thought to originate from mergers \citep{garcia-berro2012,toonen2017,Cheng2019}.

\subsection{Gravitational scattering}

Even if a planet survives tidal interactions with the star and, if applicable, in-spiral inside the envelope, the
planet may then be subject to gravitational interactions with other, more distant surviving planets. The result could
be ejection from the system, collision with the star or collision with the other planet.

Gravitational scattering between multiple major planets in giant branch planetary systems
has been investigated in only about one dozen papers (an order of magnitude smaller than 
the comparable body of literature for multi-planet scattering in main-sequence planetary systems).
Nevertheless, we know that in evolving single-star exo-systems, stellar mass loss can trigger instability due to changes in the Hill
stability limit \citep{debsig2002,veretal2013a,voyetal2013,veretal2018b}, Lagrange stability 
limit \citep{musetal2014,vergae2015,veretal2016,musetal2018} 
and shifting location of secular resonances \citep{smaetal2018,smaetal2019a}. Further, the combined effect of mass loss and
Galactic tides can incite instability from a distant massive planetary companion (like the theorized
``Planet Nine''; {\rev see \citealt*{batygin2019} for a review}) in an otherwise quiescent system \citep{veras2016b}.

The timescale over which the instability acts would be crucial for evolved $6-8M_{\odot}$ planetary systems. 
The only two of the above-cited papers which has simulated multi-planet instability in $6-8M_{\odot}$ systems
are \cite{veretal2013a} and \cite{voyetal2013}. Figures 9-12 of \cite{veretal2013a} reveal that instability 
time is unrestricted but strongly dependent
on the initial separations of the planets. Further, instability is relatively rare during the giant branch phases,
particularly for the shortest such phases at the highest stellar masses. This rarity suggests that gravitational 
instability is unlikely to feature until the star has become a white dwarf. 

A plausible scenario
is that gravitational instability amongst multiple planets (which survived the giant branch phases of evolution)
occurred a few hundred Myr after the parent star became a white dwarf. These timescales match with instability timescales for plausible sets of initial conditions from the studies above (by extrapolating from lower stellar masses). Nevertheless, multiple planets are not actually required to pollute the white dwarf with minor planets \citep{bonetal2011,debetal2012,frehan2014,antver2016,antver2019}, but can facilitate the process \citep{veretal2016,musetal2018,smaetal2018,smaetal2019a}.

\section{Minor planet survival}

The predominant focus in the exoplanet formation literature has been on major planets.
However, white dwarf metal pollution highlights the need to also model minor planet formation. In fact, almost all observations of white dwarf planetary systems are of the remnants of minor planets.

By ``minor planets'', we refer to {\rev an object with a mean radius of under $10^3$ km. This} size regime encompasses moons, comets, asteroids, and large fragments of major planets. In fact, the minor planet currently orbiting the white dwarf WD 1145+017 may best be classified as an ``active asteroid'' \citep{vanetal2015} whereas the one orbiting SDSS J1228+1040 is perhaps best described as a ferrous chunk of a fragmented planetary core \citep{manetal2019}. The third, which orbits ZTF J0139+5245 \citep{vanetal2019}, still defies classification, until at least more data is obtained.

These distinctions are becoming increasingly relevant given our detailed knowledge of white dwarf planetary systems.
A related question of interest is whether white dwarf metal pollution arises from primarily intact or
destroyed minor planets, and the locations at which those bodies or debris were scattered.
The answer depends more strongly on radiation than gravity \citep{veretal2015a} through the Yarkovsky
and YORP effects.

The Yarkovsky and YORP effects are recoil forces and torques produced due to non-uniform scattering
of absorbed radiation \citep{voketal2015}. The Yarkovsky effect changes the orbit of a spherical or 
aspherical minor planet, whereas the YORP effects spins up or/and spins down an aspherical minor planet. Both effects have been modelled for solar system asteroids in exquisite detail \citep[e.g.][]{rozgre2012,rozgre2013,cotetal2015,yuetal2018,hirsch2019}. Contrastingly,
only a handful of basic dedicated extrasolar post-main-sequence studies have been published
for the Yarkovsky effect \citep{veretal2015a,veretal2019b} and the YORP effect 
(\citealt*{veretal2014a,versch2019}).

This contrast does not suit the added complexity introduced in giant branch systems, 
when the luminosity of the parent star changes on short timescales, and can be
five orders of magnitude higher than the Sun's (see Table \ref{stelprop}). Such high luminosities easily fling
exo-asteroids out to distances of tens, hundreds or even thousands of au, regardless
of the presence of major planets, which only temporarily impede such migration 
\citep{veretal2019b}. These minor planets could also
be propelled inwards, and the direction of motion is a function of the bulk shape, spin, surface
topography and internal homogeneities of the objects.

If, however, the minor planets veer too close to a giant branch star, then they could
be spun up to disruption. This rotational fission occurs when the spin of
a minor planet reaches the critical spin failure rate $\omega_{\rm fail}$.
For strengthless rubble piles in the solar system, this rate corresponds to a
rotational period of about 2.3 hours and has robust confirmation through
observations \citep{warnetal2009,schetal2015,poletal2017}.

\cite{veretal2014a} demonstrated that asteroid belts within about 7 au
of their parent stars along the main sequence 
(including the solar system's asteroid belt) would be 
easily fragmented by YORP-induced rotational fission along the giant 
branch phases. However, they considered parent stars with main sequence masses
of $1-5M_{\odot}$. 

Now we extend their study to higher stellar masses. The spin of a minor planet
evolves according to \citet{scheeres2007,scheeres2018}

\begin{equation}
\frac{d\omega(t)}{dt}
=
\frac
{3 \mathcal{C} \Phi  }
{4 \pi \rho R^2 a(t)^2\sqrt{1 - e(t)^2} }
\left(\frac{L_{\star}(t)}{L_{\odot}}\right)
\label{YORPsimp}
\end{equation}

\noindent{}where $\rho$ and $R$ represent the minor planet's density and initial radius (both
assumed to remain constant), and $a$ and $e$ represent
its time-dependent semimajor axis and eccentricity. $\mathcal{C}$ is a constant
which represents the extent of the minor planet's asphericity, and 
$\Phi = 1 \times 10^{17}$ kg$\cdot$m/s$^2$ is the Solar radiation constant.

Equation (\ref{YORPsimp}) helps illustrate that the increasing luminosity of the
star competes against the increasing semimajor axis from stellar mass loss. As previously mentioned, for adiabatic
mass loss, the eccentricity remains constant. We simply set it to zero here.
An analytic expression of the failure rate which includes the minor planet's tensile, 
uni-axial strength $S$ is \citep{sansch2014,scheeres2018}

\begin{equation}
\omega_{\rm fail} = \sqrt{\frac{4 \pi G \rho}{3} + \frac{4S}{3 \rho R^2}}
.
\end{equation}

We now compute simple estimates of $a_{\rm break}$, the minimum initial (ZAMS) semimajor axis
for which a minor planet will survive asymptotic giant branch YORP spin-up and reach the white
dwarf phase. To do so, we integrate equation (\ref{YORPsimp}) assuming $S=0$, $e=e(t)=0$ and
an initially non-spinning minor planet. We also assume that the asteroid does not spin
down at any point during its evolution, and its asphericity is encompassed entirely within
the assumed constant value of $\mathcal{C}$. We consider the YORP effect in isolation, decoupled
from the Yarkovsky effect; such a coupling would require a significant modelling effort, and
has not yet been achieved in exoplanetary science.

\begin{figure}
\includegraphics[width=8cm]{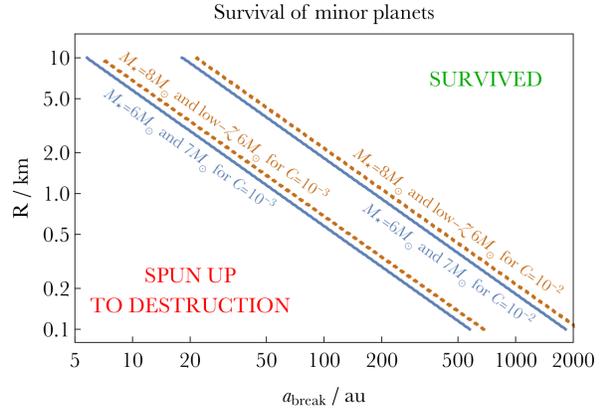}
\caption{
Minimum main-sequence semi-major axis ($a_{\rm break}$) for which a minor 
planet will not spin up to breaking point by asymptotic giant branch radiation and survive until the white dwarf 
phase. The minor planet's density is assumed to be 2 g/cm$^3$ and we apply two values for its topological
asphericity parameter ($\mathcal{C} = 10^{-2}, 10^{-3}$).  The curves for the $Z=0.02$, $M_{\star}=6M_{\odot}$
and $Z=0.02$, $M_{\star}=7M_{\odot}$ cases are visually indistinguishable. Coincidentally, so are the curves for the $Z=0.02$, $M_{\star}=8M_{\odot}$
and $Z=0.0001$, $M_{\star}=6M_{\odot}$ cases. The minor planets are assumed to be strengthless rubble piles, initially have zero
spin, and are assumed to never spin down throughout their evolution.
}
\label{minorYORP}
\end{figure}

Figure \ref{minorYORP} illustrates the result in $a_{\rm break}$--$R$ space, across a
range of $R$ (100 m -- 10 km) for which the YORP effect is significant. Coincidentally, 
the curves for the $Z=0.02$, $6M_{\odot}$ and $7M_{\odot}$ cases are visually indistinguishable
on the plot, as are the curves for the $Z=0.02$, $8M_{\odot}$ and $Z=0.0001$, $6M_{\odot}$ cases.
Changing the asphericity parameter by one order of magnitude noticeably shifts the curves.

In comparison to the survival capacity of major planets, we find that across almost the entire phase space, $a_{\rm break} > d_{\rm eng}$ and usually, $a_{\rm break} \gg d_{\rm eng}$.  However, in individual cases, these relations do not prevent the major planet from, for example, residing exterior to intact minor planets. Nevertheless, the relations suggest that a surviving major planet which is surrounded by fragmented debris is not an unreasonable scenario. 

An important related question is whether a minor planet would be sublimated from the host star before or during YORP spin-up. In order to estimate the sublimation rate of a minor planet, we use the approximation employed by \cite{jura2008}; for a more detailed description of its physical assumptions, see Section 6.1.1. of \cite{veras2016a}. The radius of a minor planet evolves according to

\begin{equation}
\frac{dR}{dt} = -\frac
{1.5 \times 10^{10} {\rm kg} \cdot {\rm m}^{-2} {\rm s}^{-1}}
{\rho}
\sqrt{\frac{T_{\rm sub}}{T(t)}} \,
{\rm exp}\left(-\frac{T_{\rm sub}}{T(t)} \right)
\label{subeq}
\end{equation}

\noindent{}where, with the the sublimation temperature $T_{\rm sub}$ and the Stefan-Boltzmann constant $\sigma$, the temperature of the minor planet $T$ can be approximated by

\begin{equation}
T(t) = \left(  
\frac
{L_{\star}(t)}
{16 \pi \sigma r(t)^2}
\right)^{1/4}
.
\end{equation}

The functional form of equation (\ref{subeq}) reveals a strong dependence on the ratio $T_{\rm sub} /T(t)$, and the value of $T_{\rm sub}$ is composition-dependent. For olivine, $T_{\rm sub} \approx 6.5 \times 10^4$ K. By adopting this value, along with $\rho = 2$ g/cm$^3$, we find that the sublimation rate is negligible across almost the entire parameter space and can be ignored.  For example, at the inner edge of this space, we find that for $r = \left\lbrace 20,15,10 \right\rbrace$ au, the \textit{maximum} sublimation rate, as measured by radius decrease, across all of our $Z=0.02$ stellar evolution tracks are, respectively, $\left\lbrace 2 \times 10^{-11}, 5 \times 10^{-8},  6 \times 10^{-4}\right\rbrace$ m/yr. Instead, for the $Z=0.0001$, $M=6M_{\odot}$ track, there is a greater danger at 10 au, with corresponding \textit{maximum} sublimation values of $\left\lbrace 3 \times 10^{-8}, 3 \times 10^{-5},  1 \times 10^{-1}\right\rbrace$ m/yr.

{\rev Given that these minor planets largely survive sublimation, one may inquire about the size distribution of the debris which results from YORP-based break-up. Although this topic requires further dedicated study, \cite{versch2019} explored this size distribution in a basic, analytical fashion, based on the results of \cite{scheeres2018}. \cite{versch2019} estimated the total number of fissioned components of an asteroid, as well as the sizes of the components, based on a number of factors including internal strength, initial spin and stellar luminosity. The result is hence dependent on many parameters and assumptions. The resulting dust produced may be blown away due to radiation pressure by comparing the interplay between radiation and gravity \citep{veretal2015a}; the presence of a major planet would affect this balance, and affects the regions where dust is ejected in a non-trivial fashion (Zotos \& Veras, In Preparation).}

\section{Discussion}

Having computed the survival limits for both major and minor planets orbiting $6M_{\odot}-8M_{\odot}$ stars, we now discuss topics which are related to these findings.

\subsection{Ice line locations} 

Giant planets are commonly assumed to have formed beyond the ice line (or ``snow line" or ``frost line")\footnote{See \cite{batetal2016} for a challenge to the canonical theory.}, where volatile compounds condense. For $1M_{\odot}$ stars, the ice line is often assumed to reside at about 2.7 au \citep{hayashi1981}, exterior to $d_{\rm eng}$.

However, as the host star mass increases, the relation between the ice line distance (denoted as $d_{\rm ice}$) and $d_{\rm eng}$ does not become immediately clear. In the ZAMS stellar mass range considered in this paper, $d_{\rm eng}$ for giant planets is between about 5.5 and 6.5 au (Fig. \ref{major}). 

In these same systems, the location of the ice line is determined by a combination of stellar irradiation and viscous heating. However, discs around massive stars are dispersed rapidly \citep{heretal2005}, allowing the effect of irradiation to dominate. In this limit, we can approximate the midplane ice-line temperature $T_{\rm ice}$ just as \cite{kenken2008} did by applying the following flared-disc prescription from
\cite{adashu1986}, \cite{kenhar1987} and \cite{chigol1997}:

\begin{equation}
T_{\rm ice} = T_{\star} \left( 
\frac{\alpha}{2}
\right)^{1/4}    
\left(
\frac{R_{\star}}{d_{\rm ice}}
\right)^{3/4}
\end{equation}

\noindent{}with

\begin{equation}
\alpha = 
0.005 \left(\frac{d_{\rm ice}}{1 \ {\rm au}} \right)^{-1}
+
0.05 \left(\frac{d_{\rm ice}}{1 \ {\rm au}} \right)^{2/7}
.
\end{equation}

By setting $T_{\rm ice} = 170$ K and ZAMS values for $R_{\star}$ and $T_{\star}$ from the aforementioned {\tt sse} code, we find that $d_{\rm ice} = 2.4-3.6$~au $\approx 0.5 d_{\rm eng}$. Hence, even in this rough approximation, the engulfment distance along the giant branch phases is more restrictive than the ice line with regard to giant planet formation location.

The theoretical considerations in this subsection help motivate dedicated planet formation studies for high host-star mass systems. We now present observational motivation.

\subsection{Future white dwarf host mass increases}

\textit{Gaia} is expected to discover, through astrometry, at least a dozen giant planets orbiting white dwarfs at the final data release \citep{peretal2014}. If realized, this exciting prediction would revolutionize our understanding of white dwarf planetary systems\footnote{Some current observational limits have been placed on the possible locations of major planets orbiting white dwarfs. Null results in a deep {\em Spitzer} survey can rule out at least $>$10\,Jupiter-mass planets within the 5-50 au range around more than 40 white dwarfs \citep{farihi2008,kilic2009}. Monitoring the stable pulsation arrival times of at least 6 white dwarfs can exclude $>$3\,Jupiter-mass planets in a range of roughly $2-5$\,au \citep{winget2015}. Unfortunately none of the most massive pulsating white dwarfs have stable, coherent modes that can be used to monitor for substellar companions.
}, but is highly unlikely to increase the mass of the most massive known white dwarf host star.

Instead, the new ``record-breaker" is more likely to arise from the steep and imminent increase in known polluted white dwarfs. This increase will arise from the larger number of new white dwarfs which have already been identified from \textit{Gaia}. \textit{Gaia} Data Release 2 \citep{gaia2018} recently uncovered an all-sky sample of $\approx$ 260,000 white dwarf candidates that is homogeneous down to $G <$ 20 mag  \citep{genetal2019}. From this sample, on the order of 10,000 white dwarfs with atmospheric planetary remnants are suitable to low-resolution spectroscopic study in the near future with wide-area multi-object spectroscopic surveys such as WEAVE \citep{weave}, 4MOST \citep{4most}, DESI \citep{desi} and SDSS-V \citep{sdss-v}.

\subsection{The smallest white dwarf hosts}

The most massive known white dwarf planetary system hosts would also represent the smallest known white dwarf planetary system
hosts. The classic mass-radius relation for white dwarfs \citep{nauenberg1972}

\begin{equation}
\frac{R_{\star}}{R_{\odot}} \approx 0.0127 \left(\frac{M_{\star}}{M_{\odot}}\right)^{-1/3}
\sqrt{1-0.607\left(\frac{M_{\star}}{M_{\odot}}\right)^{4/3} }
\end{equation}

\noindent{}yields, for the range of masses in Table \ref{stelprop}, host star radii between $0.0013R_{\odot}-0.0063R_{\odot} \approx 900$ km $-$ $4,450$ km 
$\approx 0.26R_{\rm Mars}-1.31R_{\rm Mars}$. In contrast, a typical $0.6M_{\odot}$ white dwarf gives a radius of about $2.6R_{\rm Mars}$.

How do the much smaller radii of the systems highlighted here affect the dynamical origin of metal pollution?
The Roche radii of white dwarfs scale as $M_{\star}^{1/3}$ regardless of the composition or spin of orbiting
companions \citep{veretal2017}. Therefore, doubling the white dwarf mass increases the Roche radii by about
$25$ per cent. This level of increase is significant, and can for example, trigger thermal self-disruption of passing gaseous planets which otherwise would be safe \citep{verful2019}. The larger Roche radius also provides a larger target for minor planets, which, when combined with the location requirements for major planets in these systems, places restrictions on where and when scattering  \citep{antver2016,antver2019} occurs.

Another consequence of decreasing the size of a white dwarf and increasing its Roche radius is the resulting change in structure of a debris disc formed from the disruption of a minor planet 
\citep{graetal1990,jura2003,debetal2012,veretal2014b,veretal2015b,malper2020a,malper2020b}. The outer edge would be larger than those of the currently known discs, increasing the probability of capture of solid objects \citep{griver2019}; the consequences for the inner edge is less clear because of our lack of observations between the inner disc rim and the white dwarf photosphere. The resulting disc evolution and accretion rates onto the white dwarf might also differ from what has been previously
modelled \citep{bocraf2011,rafikov2011a,rafikov2011b,metetal2012,rafgar2012,kenbro2017a,kenbro2017b,mirraf2018}.

\subsection{Other features of $6M_{\odot} - 8M_{\odot}$ planetary systems}

Major planets which form and reach distances of several to many au in under
70 Myr around $6M_{\odot}-8M_{\odot}$ stars may help identify the dominant planet formation mechanisms
in these systems. These planets formed in discs which are subject to time-varying radiative forcing
at a more extreme level than their lower-host mass counterparts. When combined
with photoevaporative effects from other stars in their birth cluster, outward migration of
planets within these discs may be enhanced \citep{verarm2004,rosetal2013,rosetal2015,ercros2015,jenetal2018}.

If the $6M_{\odot}-8M_{\odot}$ progenitor host stars for polluted white dwarfs were slightly more massive, then they would have instead become neutron stars; {\rev planets are known to exist around single pulsars} \citep{wolfra1992,wolszczan1994,konwol2003}. These pulsar planets most likely represent products of second-generation planet formation from mergers or fallback \citep{curhan2007,marmet2017} although questions still remain \citep{hanetal2009,keretal2015}\footnote{The sudden stellar mass loss experienced in a supernova is typically insufficient to eject a planet on a compact orbit \citep{veretal2011}. The more pertinent but outstanding question of first-generation planet survival is whether the planet can physically withstand a supernova's stellar ejecta.}. Exo-asteroids may orbit pulsars below the detection threshold \citep{corsha2008,shaetal2013,daietal2016,smaetal2019b}, although that threshold already lies at about 1 per cent of the mass of the Moon \citep{behetal2019}.

In contrast, both major and minor planets orbiting single white dwarfs are first-generation, because the mass in the white dwarf debris discs required to form new planets is at the level of an Io mass or higher \citep{vanetal2018}. In contrast, the masses of white dwarfs with observed debris discs remain uncertain but are assumed \citep{vanetal2015,vanetal2019} to be generated instead by asteroids (with a typical mass two orders of magnitude less than Io). Further, the chemistry of any second-generation planets would be recycled from broken-up first generation bodies; Veras, Karakas \& G\"{a}nsicke (submitted) demonstrate definitively that white dwarf pollution does not arise from stellar fallback.

\section{Summary}

The most massive currently known progenitor of a white dwarf planetary system host ($\approx 4M_{\odot}$) exceeds the maximum mass of a main-sequence major planet host ($\approx 3M_{\odot}$). Because this mass difference is likely to increase with white dwarf population growth (Section 5.2), we have investigated major and minor planet survival limits for the highest possible stellar masses ($\approx~6M_{\odot} - 8M_{\odot}$) which yield white dwarfs. This mass regime represents almost completely unexplored territory for planet formation theorists, and largely unexplored territory for post-main-sequence planetary evolution investigators.

We found that a major planet needs to reside beyond 3-6 au (depending on the type of planet) at the end of the short (40-70 Myr) main-sequence phase in order to survive its host star transition into a white dwarf. Minor planets could have remained intact
or broken up due to YORP-induced giant branch radiation, having been subject to host star luminosities reaching up to $10^5L_{\odot}$. The boundary separating these two possibilities ($\sim 10^1-10^3$ au) is very sensitive to minor planet size, but is high enough to suggest that metal pollution in these systems would originate from already fragmented debris rather than intact minor planets which fragment upon reaching the white dwarf Roche radius. Our study motivates dedicated planet formation studies around the highest mass stars, and predicts that metal pollution can occur in the highest mass white dwarfs.

\section*{Acknowledgements}

{\rev We thank the expert reviewer Fred C. Adams for helpful comments which have improved the manuscript}.
DV gratefully acknowledges the support of the STFC via an Ernest Rutherford Fellowship (grant ST/P003850/1). The research leading to these results has received funding from the European Research Council under the European Union's Horizon 2020 research and innovation programme no. 677706 (WD3D). Support for this work was also provided by NASA through grant number HST-GO-15073.005-A from the Space Telescope Science Institute, which is operated by AURA, Inc., under NASA contract NAS 5-26555. FM acknowledges support from the Royal Society
Dorothy Hodgkin Fellowship. GMK is supported
by the Royal Society as a Royal Society University Research Fellow. BTG is supported by the UK Science
and Technology Facilities Council grant ST/P000495.

\label{lastpage}
\end{document}